\documentclass[a4paper,11pt]{article}
\usepackage{aaskaiid}
\usepackage{orcidlink}
\usepackage{aas_macros}
\usepackage{gensymb}
\usepackage{amsmath,siunitx}
\usepackage{xeCJK}

\title{Demographics of planet-forming disks with the SKAO}
\ShortTitle{Demographics of planet-forming disks}

\author[1]{Antonio Garufi\orcidlink{0000-0002-4266-0643}}
\ShortName{Antonio Garufi et al.} 
\author[2,3,4]{Sebasti\'an P\'erez\orcidlink{0000-0003-2953-755X}}
\author[5]{John D.\ Ilee\orcidlink{0000-0003-1008-1142}}
\author[6,7]{Daniel J. Price\orcidlink{0000-0002-4716-4235}}
\author[8]{\mbox{Paola Pinilla}\orcidlink{0000-0002-3123-1544}} 
\author[7]{Marion Villenave\orcidlink{0000-0002-8962-448X}}
\author[9]{Eleonora Bianchi\orcidlink{0000-0001-9249-7082}}
\author[10]{Luca Cacciapuoti\orcidlink{0000-0001-8266-0894}}
\author[7]{\mbox{Greta Guidi}\orcidlink{0000-0002-7002-8928}}
\author[9]{Giovanni Sabatini\orcidlink{0000-0002-6428-9806}}
\author[11,12]{Yinhao Wu (吴寅昊)\orcidlink{0000-0003-3728-8231}}
\author[13]{Asmita Bhandare\orcidlink{0000-0002-1197-3946}}
\author[9]{Claudio Codella\orcidlink{0000-0003-1514-3074}}
\author[7]{Nicol\'as Cuello\orcidlink{0000-0003-3713-8073}}
\author[14]{Liton Majumdar\orcidlink{0000-0001-7031-8039}}
\author[15]{Mayank Narang\orcidlink{0000-0002-0554-1151}}
\author[9]{\mbox{Linda Podio}\orcidlink{0000-0003-2733-5372}}
\author[16]{Danae Polychroni\orcidlink{}}
\author[5]{Isaac Radley\orcidlink{0009-0007-2837-8207}}
\author[17]{Jessica Speedie\orcidlink{0000-0003-3430-3889}}
\author[18]{\mbox{Leonardo Testi}\orcidlink{0000-0003-1859-3070}}
\author[19, 20, 9]{Claudia Toci\orcidlink{0000-0002-6958-4986}}
\author[16]{Diego Turrini\orcidlink{0000-0002-1923-7740}}
\author[21]{David Wilner\orcidlink{0000-0003-1526-7587}}

\affiliation[1]{INAF, Istituto di Radioastronomia, Via Gobetti 101, I-40129, Bologna, Italy}
\emailAdd{antonio.garufi@inaf.it}
\affiliation[2]{Departamento de F\'isica, Universidad de Santiago de Chile, Av.\,Victor
Jara 3493, Santiago, Chile}
\affiliation[3]{Millennium Nucleus on Young Exoplanets and their Moons (YEMS), Chile}
\affiliation[4]{Center for Interdisciplinary Research in Astrophysics Space Science (CIRAS), Universidad de Santiago, Chile}
\affiliation[5]{School of Physics and Astronomy, University of Leeds, Leeds, UK, LS2 9JT, UK}
\affiliation[6]{School of Physics and Astronomy, Monash University, Clayton VIC 3800, Australia}
\affiliation[7]{Univ. Grenoble Alpes, CNRS, IPAG, 38000 Grenoble, France}
\affiliation[8]{Mullard Space Science Laboratory, University College London, Holmbury St Mary, Dorking, Surrey RH5 6NT, UK}
\affiliation[9]{INAF, Osservatorio Astrofisico di Arcetri, Largo E. Fermi 5, I-50125, Firenze, Italy}
\affiliation[10]{European Southern Observatory, Alonso de Cordova 3107, Vitacura, Region Metropolitana de Santiago, Chile}
\affiliation[11]{Shanghai Astronomical Observatory, Chinese Academy of Sciences, Shanghai 200030, People's Republic of China}
\affiliation[12]{School of Physics and Astronomy, University of Leicester, Leicester LE1 7RH, UK}
\affiliation[13]{Universit\"{a}ts-Sternwarte, Fakult\"{a}t f\"{u}r Physik, Ludwig-Maximilians-Universit\"{a}t M\"{u}nchen, Scheinerstr. 1, 81679 M\"{u}nchen,Germany}
\affiliation[14]{Exoplanets and Planetary Formation Group, School of Earth and Planetary Sciences, National Institute of Science Education and
Research, Jatni 752050, Odisha, India}
\affiliation[15]{Jet Propulsion Laboratory, California Institute of Technology, 4800 Oak Grove Drive, Pasadena, CA 91109, USA}
\affiliation[16]{INAF - Turin Astrophysical Observatory, Via Osservatorio 20, I-10025, Pino Torinese, Italy}
\affiliation[17]{Department of Physics \& Astronomy, University of Victoria, Victoria, BC V8P 5C2, Canada}
\affiliation[18]{Alma Mater Studiorum Universit\`a di Bologna, Dipartimento di Fisica e Astronomia (DIFA), Via Gobetti 93/2, 40129 Bologna, Italy}
\affiliation[19]{Departamento de Fisica aplicada III, ETSI Universidad de Sevilla,  Camino de los Descubrimientos, 41092 Sevilla}
\affiliation[20]{European Southern Observatory, Karl-Schwarzschild-Strasse 2, D-85748 Garching bei München, Germany}
\affiliation[21]{Center for Astrophysics $\vert$ Harvard \& Smithsonian, 60 Garden St., Cambridge, MA 02138, USA}

\abstract{Understanding how solid material in planet-forming disks evolves from \mbox{micron-sized} dust to planetary cores is a central challenge in modern astrophysics. This study has advanced dramatically in the past decade, largely driven by ALMA and high-contrast imaging facilities. However, major uncertainties remain regarding the presence, evolution, and role of centimeter-sized grains (the pebbles) in planet formation. The SKAO will fill this gap by enabling the first large-scale, high-resolution survey of disk emission at centimeter wavelengths. This chapter presents the scientific rationale and observational strategies to detect and characterize pebbles in the planet-forming disks of nearby star-forming regions. By resolving their spatial distribution, spectral properties, and evolutionary trends, SKA will offer essential constraints on dust growth and disk dynamics. This work provides observational strategies, target selection, and predictions on the detectability of hundreds of nearby disks. The chapter also explores SKA's potential to uncover the actual dust mass in disks, protoplanets and their circumplanetary disks, and other aspects of the planet formation. Together, these capabilities will establish SKAO as a cornerstone facility for planet formation science in the coming decade.}

\begin{document}

\maketitle

\section{Introduction} \label{sec:intro}
The more than {8,100} exoplanets discovered to date\footnote{The Extrasolar Planets Encyclopaedia at \url{https://exoplanet.eu}.} showcase an extraordinary variety in both system architecture and individual planetary characteristics -- from tightly packed super-Earth systems to giant planets on eccentric orbits. The high occurrence rate of exoplanets \citep[{of the order of 30--50\% for close-in planets around solar-type stars}, see e.g.][]{Fressin2013, Fulton2017} highlights the efficiency of planet formation. At the same time, the striking diversity of planetary systems suggests that planet formation unfolds under a wide range of initial conditions. The study of the earliest stages of planet formation -- spanning their initial assembly from interstellar grains to their early interactions with each other and the surrounding natal material -- has become a central focus in modern astrophysics. This research has made significant progress over the past decade, largely due to advancements in optical and near-IR high-contrast imaging \citep[see review by][]{Benisty2023} and the capabilities of the Atacama Large Millimeter/submillimeter Array \citep[ALMA, see reviews by][]{Andrews2020, Bae2023}.

Circumstellar disks form as a natural by-product of the star formation process within molecular clouds -- cold, dense, magnetized regions of interstellar material (composed of approximately 99\% gas and 1\% dust). The turbulent collapse of clouds produces multiple star systems, while the residual angular momentum of infalling gas leads to the formation of circumstellar and circumbinary disks \citep{Offner2023}. This initial phase corresponds to the Class~0 and Class~I stages of young stellar object (YSO) evolution. During these phases, accretion onto central protostars are accompanied by powerful outflows, including wide-angle disk winds and highly collimated jets, which help remove excess angular momentum \citep{Zhao2020}. Once the surrounding envelope has dissipated, the system transitions into the Class II phase, during which the disk (now largely unobscured) is directly observable over a wide range of wavelengths. These processes unfold in complex environments shaped by multiple stellar encounters in crowded star-forming regions \citep{Cuello2023}, as well as by dynamical interactions with the surrounding medium in the form of inflows, filaments, and continuing accretion from the ambient cloud \citep{Pineda2023}.

The vast majority of stars still surrounded by a disk are found in star-forming regions. The nearest of these regions --- such as Corona Australis, Chamaeleon, and the Scorpius-Centaurus association --- are situated between 100 and 200 pc from Earth. This means that a circumstellar disk spanning several tens of astronomical units would have an angular size of only a fraction of an arcsecond. Consequently, resolved observations of these disks became routinely possible only with the advent of ALMA and of the extreme adaptive optics imagers mounted on 8-m class telescopes. Before these observations, spatially resolved maps were available for a handful of extraordinarily bright and extended disks with the Hubble Space Telescope (HST) and previous generation instrumentation \citep[e.g.][]{Pantin2000, Mouillet2001, Duchene2004, Pietu2007, Brown2009} and also with previous generation interferometers such as SMA \citep[e.g.][]{Andrews2005}.  

The results of the first ALMA Long Baseline Campaign in 2014 proved spectacular. These revealed `an astonishing level of detail in the circumstellar disk' surrounding the young star HL Tau, consisting of a series of alternating rings and gaps in the continuum emission \citep{ALMA2015}. This result revolutionised the field of planet formation, almost overnight changing it from a mostly theoretical field focussed on the origins of our own solar system, to the data-driven, observational field it has been ever since. Prior to the ALMA observations, HST images of \mbox{HL Tau's} surroundings showed collimated jets of gas emanating from the disk, typical of stars in the earliest stages of formation. The problem in HL Tau is that if the gaps were caused by planets \citep[e.g.][]{Dipierro2015,Jin2016}, the young age (less than 1 Myr) causes problems for our ideas about planet formation. This led to intense speculation that the gaps were not caused by planets but by other physics (such as condensation fronts, see e.g. \citealt{Zhang2015}, or gas-dust instabilities, see review by \citealt{Lesur2023}).  

{Whether the observed sub-structures of young disks are due to planets or not, the standard core accretion model \citep{Pollack1996} faces challenges in explaining the formation of gas giants even within the typical lifetimes of planet-forming disks (a few Myr). In this model, planets form through the gradual accumulation of solid material into km-sized planetesimals, which then merge to form planetary embryos (typically with masses $\sim 10^{-2}$--$10^{-1}\,M_\oplus$). These embryos continue to grow through planetesimal accretion until a core of approximately $10\,M_\oplus$ is reached, at which point the planet begins to accrete a gaseous envelope. However, this process is too slow, as planetesimal collisions are inefficient, and core growth may not reach the necessary mass before the disk dissipates.} 

\begin{figure}[h]
    \centering
    \includegraphics[width=\columnwidth]{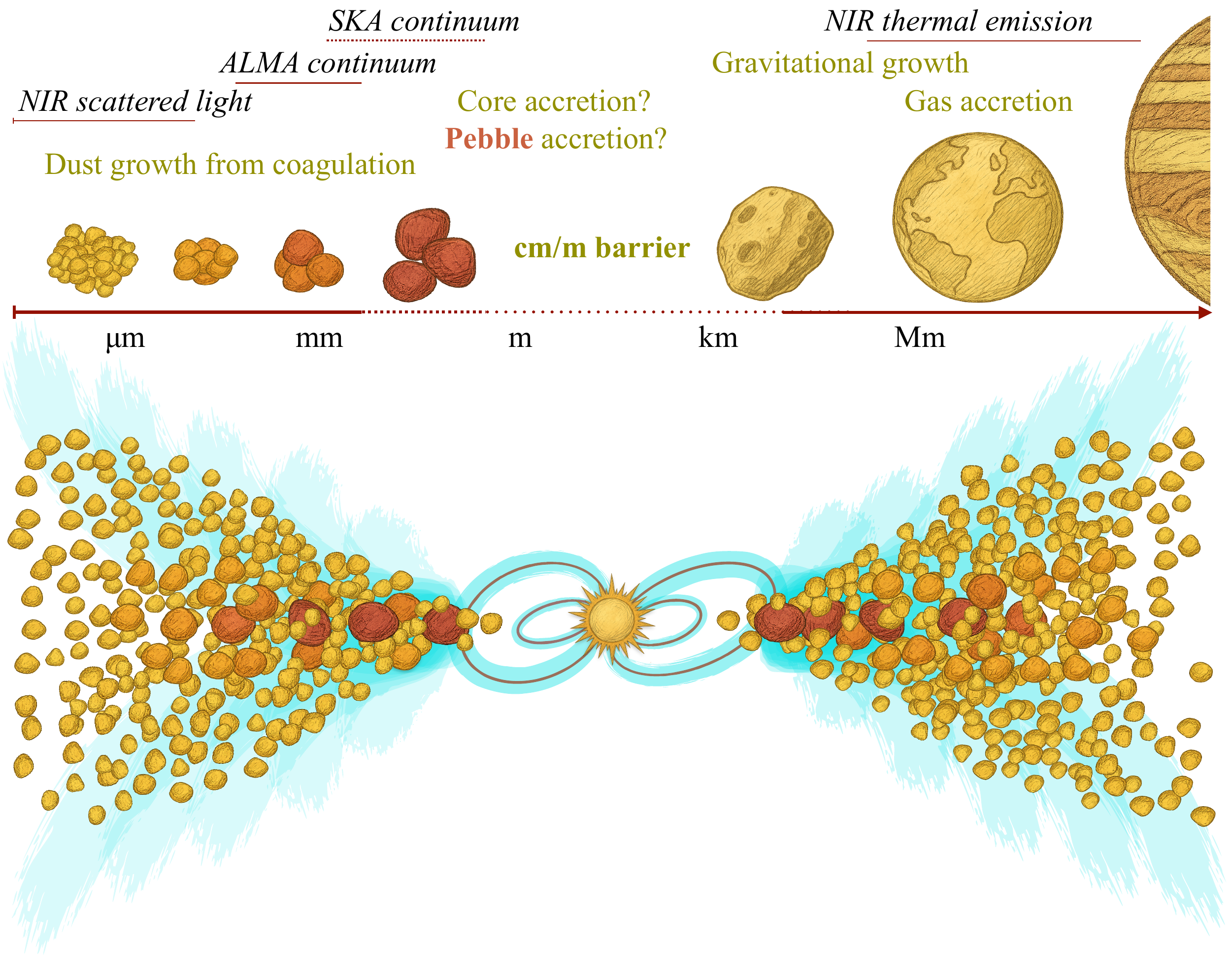}
    \caption{Sketch of the dust grain growth and disk structure. The physical processes, observational techniques, and disk components depicted are described in Sect.\,\ref{sec:intro}. {The horizontal axis in the upper panel represents increasing particle size (and implicitly evolutionary time), from micron-sized grains to planetary cores. The disk in the sketch coarsely extends to 30--100 au.} The inner part highlights the motion of the ionized gas {at (sub-)au scale} toward the star following the magnetic field.}
    \label{fig:sketch}
\end{figure}

{A more recent alternative, pebble accretion, offers a more efficient pathway for planet growth  \citep{Ormel2010, Lambrechts2012}. In this model, planetary embryos, formed at early stages (e.g. via coagulation or streaming instability), grow by accreting millimeter-to-meter-sized pebbles \citep{Johansen2019, Lyra2023}. Unlike planetesimal accretion, pebble accretion is highly efficient due to gas drag, which slows down pebbles and allows them to be captured by a growing core rather than simply bouncing off or missing entirely. This process allows planets to grow much faster, making it more feasible for gas giants to reach the same critical core mass before the disk disperses. Once this mass is reached, gas accretion proceeds as in the classical scenario, but on significantly shorter timescales due to the more rapid core growth. Figure \ref{fig:sketch} sketches the gradual growth of solids following these processes.}


Since pebble accretion relies on the efficient growth and retention of cm- and m-sized particles, confirming their existence and constraining spatial distribution in the disk is crucial for validating the theory. Generally speaking, the retention and further growth of a cm-sized particle are also puzzling since particles reaching this dimension experience a rapid inward drift due to gas drag causing them to spiral into the central star on timescales much shorter than what is needed for them to further coalesce into planetesimals \citep{Weidenschilling1977}. Additionally, collisions between these large grains tend to be destructive rather than leading to further growth, as relative velocities become high enough to cause fragmentation rather than sticking \citep{Windmark2012}. Both these effects concur to define the centimeter/meter barrier of planet-forming disks (see sketch of Fig.\,\ref{fig:sketch}). To overcome this barrier, mechanisms such as turbulent concentration and pressure bumps due to instabilities in the disk, zonal flows, or interactions with forming planets are therefore crucial. 

\begin{figure}[tp]
    \centering
    \includegraphics[width=1.0\columnwidth]{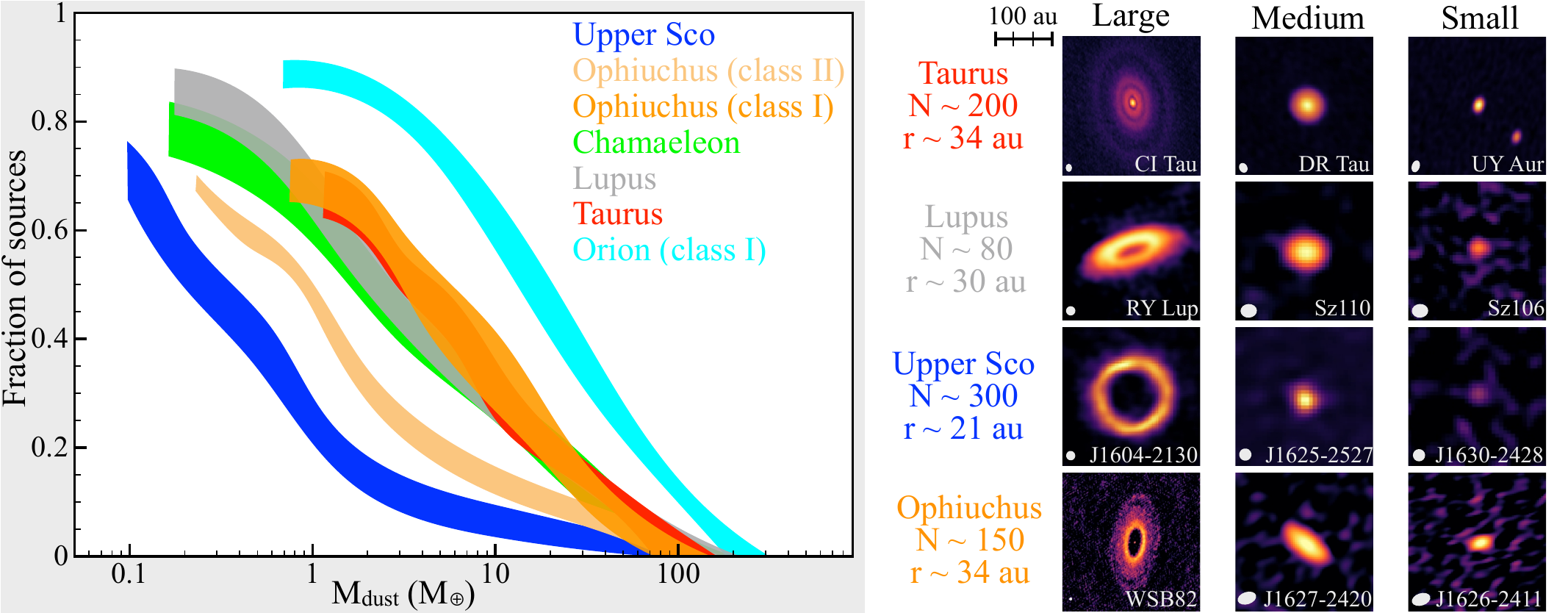}
    \caption{Disk demographics from ALMA. The left panel highlights the observed decrease of dust mass as YSOs transition from Class I to Class II, and as we observe older star-forming regions such as Upper Sco (older than 5 Myr -- in contrast to Ophiuchus, Chamaleon, Lupus, and Taurus of 1--3 Myr). To the right, an illustrative gallery of disks with different size from various star-forming regions is shown. Very extended disks such as those of CI Tau, RY Lup, J1604-2130, and WSB82 represent an exception. The {number of Class II sources and the} median disk size of the regions are given. All disks are shown with the same angular scale, and their approximate physical scale (mildly varying with the distance) is shown by the bar at the top.}
    \label{fig:disk_demographics}
    \end{figure}

The systematic detection of disk sub-structures following the HL Tau revolution \citep[see e.g.][]{Garufi2018,Andrews2020,Benisty2023} has reinforced the idea that pebbles can be trapped and concentrated in specific disk regions, helping to overcome the radial drift (or `meter-sized') barrier. Beyond HL Tau, the number of disks observed by ALMA in the past two decades has been exponentially increasing, enabling  statistically significant demographic studies of the nearby star forming regions. Such studies have highlighted a trend of decreasing dust median mass (see Sect.\,\ref{sec:opacity}) with age of the star forming region \citep[e.g.][]{Ansdell2017, Cieza2018}, as expected from the progressive aggregation of dust grains into larger bodies, undetectable at \mbox{(sub-)mm} wavelengths. This trend is shown in Fig.\,\ref{fig:disk_demographics}. In addition, a steeper correlation between M$_{\rm dust}$ and M$_*$ for older regions has been interpreted as either evidence of larger grains in older systems, or a more efficient radial drift around lower mass stars \citep[e.g.][]{Barenfeld2016, Pascucci2016, pinilla2017b}. While ALMA first prioritized observations of the brightest disks, for practical reasons, a large sample of disks with extent of tens to more than a hundred au has been subsequently gathered \citep[e.g.][]{Andrews2018, Long2018, Cieza2019}. However, the majority of disks in ALMA survey were spatially unresolved, indicating a disk size $\lesssim$ 30 au \citep[e.g.][see Fig.\,\ref{fig:disk_demographics}]{Barenfeld2016, Ansdell2016, GuerraAlvarado2024}. One main bias in these studies is that they target nearby low-mass star-forming region, with a relatively low UV irradiation. {This bias is important because the majority of stars, including the Sun, are thought to form in clustered environments exposed to strong external UV radiation. These environments can alter disk evolution through external photoevaporation, potentially affecting disk lifetimes, sizes, and ultimately planet formation pathways.} ALMA also probed $\sigma$ Orionis \citep{Ansdell2017, Huang2024}, that is a cluster that is representative of regions embedded in a strong UV field from O stars. This work showed systematically smaller disks possibly due to the effect of external photoevaporation.

One main limitation of the ALMA results obtained over the last decade is that ALMA primarily probes dust populations emitting (sub-)millimeter emission, that corresponds to grain sizes of the order of $\lambda/2\pi \sim 100-200\,\mu$m. Already at these wavelengths we find dust disks that are completely different to the gas disks in terms of radial extent and sub-structures. A key prediction of the pebble accretion hypothesis is that larger dust grains should be even more radially concentrated (see Fig.\,\ref{fig:sketch}), making the spatial extent of pebble-induced emission very challenging for the current generation of radio telescopes. This has driven further observational efforts to constrain the size distribution and spatial distribution of pebbles across different disk environments. In this way, pebble accretion has shifted the focus of disk studies toward understanding how and where cm-sized grains accumulate, survive, and participate in planet formation. 

A second severe limitation toward the characterization of dust grains from ALMA observations is the possibility that a significant fraction of the disk emission is optically thick in the (sub-)millimeter regime \citep{Ribas2020,Sierra2020}. This is particularly true for small disks or for the inner regions {($\lesssim30$ au)} of extended disks that typically exhibit a smooth and bright appearance \citep[see, e.g.][]{Andrews2018, Long2018}. The observed correlation between disk size and luminosity at millimeter wavelengths may be explained by optically thick emission \citep{Tripathi2017,Tazzari2021a}. Furthermore, dust scattering can suppress disk emission, leading to an underestimation of the optical depth \citep{Birnstiel2018, Zhu2019, Sierra2020}. 

All these results make it evident to the scientific community that (sub-)centimeter observations (5-50 GHz) are invaluable for investigating dust density and grain characteristics in disks. At these frequencies, optical depth remains relatively low, yet disk emission is likely still detectable within reasonable telescope observation times. However, to date, a (marginally) resolved map at these frequencies has only been achieved for the brightest and most extended disks with the VLA \citep[see e.g.][]{Wilner1996, Wilner2000, Wilner2005, Lommen2009, Macias2016, Carrasco-Gonzalez2016, Carrasco-Gonzalez2019, Guidi2022} or with the ATCA \citep{Casassus2015, Norfolk2021}.



This chapter {builds directly on the framework outlined by \citet{Testi2015} and \citet{Hoare2015} describing the role of centimeter interferometry in studying grain growth and disk evolution}, and outlines the scientific opportunities that the SKAO offers for studying the global morphology and the demographic distribution of the centimeter emission from planet-forming disks. In the next chapter \citep{Wu2026}, the scientific potential of the SKA for investigating the nature of disk sub-structures at these wavelengths will be discussed. Finally, the chapter by \citet{Guidi2026} will explore the ability by SKA to separate disk dust emission and ionized gas emission, and discuss the opportunity to study the latter in detail.

\section{Scientific goals and methodology}

In this section, we outline the main scientific objectives that the community can pursue with SKAO for characterizing planet-forming disks and discuss the methods for achieving meaningful results. These objectives include the detection and characterization of pebbles (Sect.\,\ref{sec:pebbles}), an improved constraint of the dust opacity and mass of disks (Sect.\,\ref{sec:opacity}), the ambitious (in)direct detections of protoplanets and circumplanetary disks (Sect.\,\ref{sec:cdp}), a better knowledge of the anomalous microwave emission (Sect.\,\ref{sec:ame}), and the determination and portrayal of eruptive stars (Sect.\,\ref{sec:eruptive_stars}).

\subsection{Pebbles in planet-forming disks} \label{sec:pebbles}

A key objective of the planet formation community is to leverage the SKAO to detect and characterize dust grains in disks that are approaching centimeter sizes. To achieve this, we will utilize the potential of SKA-Mid Band 5b observations. As detailed in Sect.\,\ref{sec:sample}, we predict that the dust emission at this frequency will be strong enough to allow detection in a significantly large number of disks within a reasonably short telescope time. In contrast, the spectral properties of the expected dust emission from disks are such that the detection of dust emission is highly challenging already in Band 5a. Nonetheless, Band 5a and lower-frequency bands will be valuable for characterizing the ionized gas of the inner disk regions (see Sect.\,\ref{sec:sample} and the chapter by \citet{Guidi2026}. In this section, we discuss three main avenues for investigating dust growth in planet-forming disks with SKAO \mbox{Band 5b} observations: the spatial and the spectral characterization of pebbles, and the full exploitation of the southern sky to portray the demography of pebbles from different regions, and thus their evolution in time. 

\subsubsection{Spatial distribution of pebbles} \label{sec:spatial_pebbles}

The first avenue is particularly compelling because {SKA-Mid AA4 configuration, with baselines extending to $\sim$150 km} will be the first instrument capable of providing spatially resolved maps of the centimeter-wavelength emission from planet-forming disks. Previous studies with the VLA and the ATCA have laid the groundwork by detecting this emission. \citet{Wilner2005} detected a marginally resolved 3.5~cm emission from the nearby young star TW Hya, and interpreted it as thermal emission from dust particles {due to its constancy} in time over weeks and years. \citet{Lommen2009} used similar arguments to claim the detection of pebbles around WW Cha, while other mechanisms were considered dominant at cm wavelengths in other sources. More recent papers \citep[e.g.][]{Perez2015, Wright2015, Carrasco-Gonzalez2016, Macias2018, Guidi2022} spatially resolved and characterized some particularly extended disks {(typically $\gtrsim 0.5''$, corresponding to $\gtrsim 75$ au at 150 pc)} from the emission up to 1--3 cm, while the unresolved longer-wavelength emission was mainly ascribed to the ionized gas. 

The best resolution that can be obtained by VLA in Band Ku (2 cm) for an object at 150 pc is as much as 30 au whereas most disks are expected to be smaller than that (see Fig.\,\ref{fig:disk_demographics}). On the contrary, the superior resolution of the {SKA-Mid (in principle up to 5 au at 3 cm in regions such as Scorpius-Centaurus at $\sim150$ pc)} will allow us to disentangle the dust emission from the ionized gas emission, and to confront the observed pebble distribution with that of the smaller grains probed by ALMA in three disk spatial extents (radially, vertically, and azimuthally/locally). First, larger dust particles that are traced at longer wavelength are expected to experience a more efficient radial drift to regions of high pressure. In the absence of pressure bumps that can trap particles, dust disk radii are expected to be smaller when observed at longer wavelengths. Thus, determining the disk radii at different wavelengths can help constrain key disk properties, such as the gas mass distribution \citep{powell2019, Franceschi2022}. In the presence of dust traps, seen as substructures in the dust continuum emission, the larger grains are expected to be more radially concentrated in the pressure maximum \citep[see chapter by][]{Wu2026}, which can inform us on the radial turbulence level and dust-to-gas coupling in the trap \citep[e.g.][]{Dullemond_2018}. In addition, the radial segregation expected between small grains and large grains can help constrain the origin of the observed structures. For example, for disks with large cavities, a different behaviour of radial segregation is expected if the cavity is formed by planet(s) or regions of low ionization known as dead zones \citep[e.g.][]{pinilla2015, pinilla2019}, with the peak of the emission shifting inwards at longer wavelengths in the case of dead zones, which is opposite to the case of planets. When a cavity is carved by a planet, the expected radial segregation depends on the level of disk turbulence and dust diffusion \citep{Ovelar2016}, which are of great importance to constrain gas and dust evolution in planet-forming disks.

Secondly, the vertical concentration of pebbles is crucial to understand how fast can pebble accretion proceed to form large planetary cores~\citep{Lambrechts2012}. Estimating the vertical extent of disks relies on several techniques requiring spatially resolved observations. In disks at intermediate inclinations, the depth of any possible disk gap depends on the vertical thickness of the surrounding rings. This has been used in a few tens of systems to infer that the vertical extent of grains visible with ALMA at about 1 mm in the outer regions of disks is typically much below $H_d/r < 0.04$~\citep[][]{Pinte2016, Pizzati_2023, Villenave_2022, Villenave_2025}, {where $H_d$ is the dust scale height and $r$ is the radial distance from the star}. Azimuthal brightness variations of optically thin and radially narrow rings also provides constraints to the disk thickness. This technique revealed that a few disks have spatially varying vertical extent, with a thicker inner ring and a thinner outer ring \citep{Doi_Kataoka_2021, Doi_Kataoka_2023, Villenave_2025}. Azimuthal brightness variations of optically thick disk regions can also reveal a bright wall, which favours the modelling of the vertical extent of some specific disk regions~\citep{Ribas_2024, GuerraAlvarado2024}. Alternatively, edge-on disks are also very favourable targets to estimate their vertical extent as it is directly exposed to the observer. Using radiative transfer modelling to reproduce the major and minor axis profile of such disks it is possible to infer their true dust scale height~\citep[e.g.][]{Villenave2020, Lin_2023}.

Larger dust particles are expected to be more vertically concentrated to the disk midplane (see sketch in Fig.\,\ref{fig:sketch}) and can thus allow for faster planet formation. Moreover, because the vertical extent of dust particles depends on the ratio between the turbulence strength, $\alpha$, and the dust-gas coupling, or Stokes number, \emph{St} {\citep{Dubrulle1995, Youdin2007}}, obtaining constraints at multiple wavelengths, up to the centimeter regime, will permit to constrain both parameters more precisely. A better knowledge of the Stokes number will also be particularly interesting to constrain the total disk mass (see Sect.\,\ref{sec:opacity}). Finally, the variation of dust vertical thickness with wavelengths can provide insight towards the physical origin of the turbulence at play (e.g.\ ideal MHD, \citealt{Fromang_Nelson_2009}, non ideal MHD, \citealt{Riols_2018}) as different mechanisms predict different dependencies with the particle size.

Finally, local (both azimuthal and radial) concentrations of pebbles to be resolved by SKAO represent the most promising feature to study pressure bumps, dust traps, and the mechanism through which pebbles overcome the aforementioned meter barrier. This topic is the main subject of the chapter by \citet{Wu2026}.

\subsubsection{Spectral properties of pebbles} \label{sec:spectral_pebbles}

The second key approach is the spectral characterization of the dust emission. While {the SKA wavelength coverage of the detectable dust emission is limited to the two points at either end of Band 5b}, we can leverage previous observations from ALMA and possibly other (sub-)mm interferometers to derive (spatially resolved) spectral indices from 0.85 mm to 3 cm. The (sub-)mm spectral indices of planet-forming disks have been routinely investigated both in the pre-ALMA \citep[e.g.][]{Andrews2005, Rodmann2006,Ricci2010,Ubach2012} and ALMA era \citep[e.g.][]{Akeson2014, Ansdell2018, Villenave2020, Tazzari2021b, Han2025}. Most of the observed disks exhibit relatively low spectral indices ($\alpha \sim 2-3$), which have been widely interpreted as indicative of a grain population with sizes extending up to the millimeter range. However, the millimeter emission from these disks is likely to be at least partially optically thick, which can artificially reduce the measured spectral indices. As a result, these measurements may not directly reflect the intrinsic dust grain properties.

\begin{figure}[h]
    \centering
    \includegraphics[width=0.75\columnwidth]{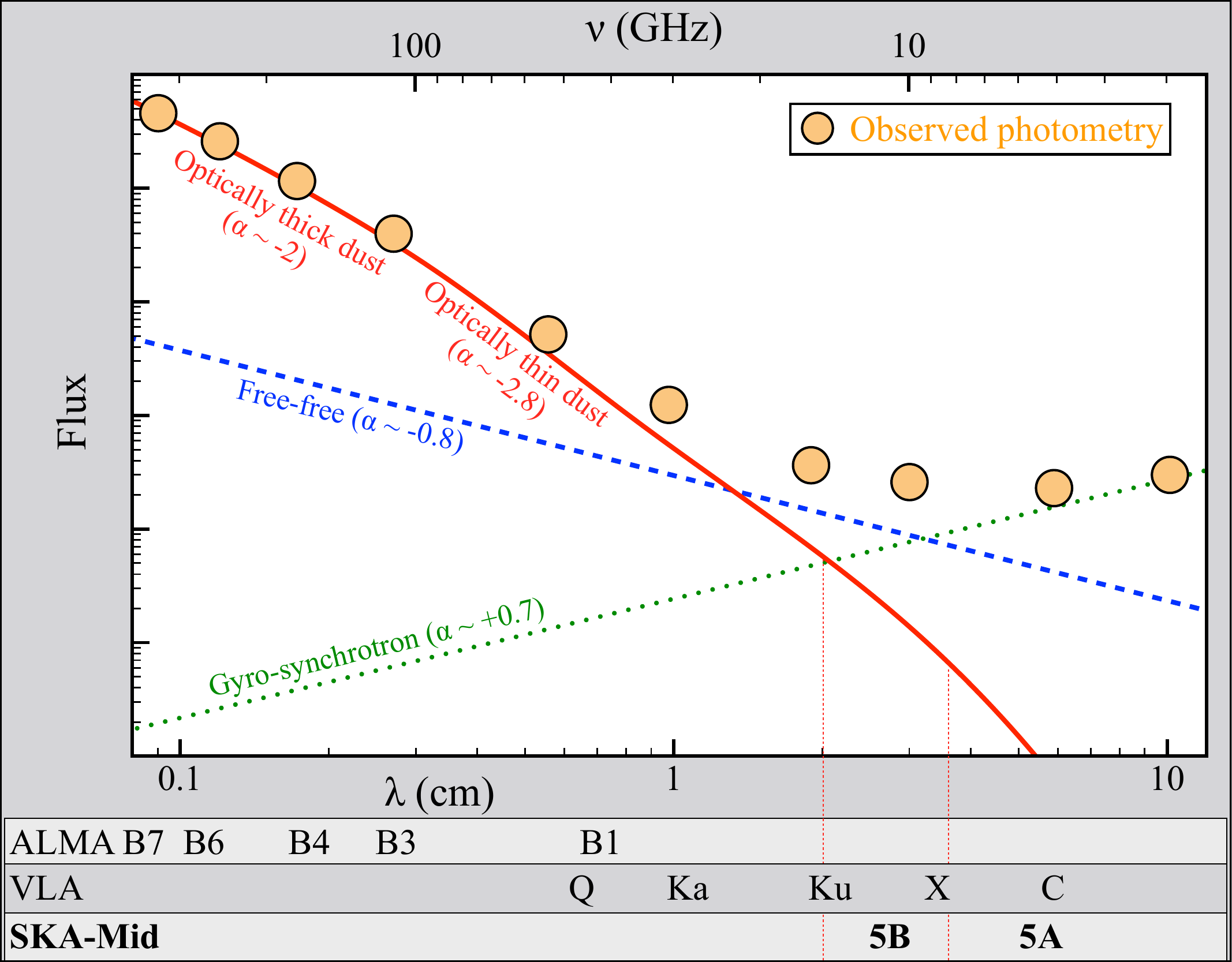}
    \caption{Sketch of SED from a young stellar object. The dust emission is optically thick up to 3 mm. After 1 cm, the contribution from the free-free emission surpasses that from the dust. The result is an abrupt flattening for the observed photometry with wavelength. In case of gyro-synchrotron emission, the trend  becomes positive after a few cm. With SKA in Band 5b, we expect the ionized gas contribution to be dominant over the dust emission although the latter would still be detectable (see Sect.\,\ref{sec:sample}).}
    \label{fig:sed_component}
\end{figure}

Further evidence that the measured spectral indices may be partially influenced by optically thick emission comes from a recent VLA survey of disks in Taurus \citep{Garufi2025}. This study employed a two-component model to disentangle the contributions of free-free and dust emission in a large set of unresolved data. The model revealed a significant shift in the median spectral index between millimeter and centimeter wavelengths (2.3 vs.\ 2.8), which was attributed to the transition to optically thin emission (as depicted in the illustrative SED of Fig.\,\ref{fig:sed_component}). According to their model, a median 35\% of the 2-cm emission would originate from the dust (see also \citealt{Guidi2022}, \citealt{Coutens2019}, and the chapter by \citet{Guidi2026}, and the grain population from the inner disk regions {($\lesssim30$ au)} responsible for the cm-emission would not significantly differ from disks with very different morphology depicted by ALMA (e.g.\ compact smooth disk vs. extended sub-structured disks). These results can be confirmed or refuted only by means of resolved maps of the spectral index.

\subsubsection{Evolution and demographics of pebbles}
The third avenue for the study of pebbles is the characterization of their evolution at different ages and under different conditions. SKAO’s location in the southern hemisphere presents the major advantage to access the most significant nearby star-forming regions (e.g.\ Chamaeleon, Lupus, Ophiuchus, Corona Australis), enabling observations of a diverse sample of objects across different environments and evolutionary stages. The clustered nature of star-forming regions also offers an opportunity with the SKAO given its wide field of view, since many disks could potentially be observed at high resolution with a single telescope pointing (see Sect.\,\ref{sec:sample_deepfield}). 

The existence of several young, protostellar systems (Class 0 and Class I) in these regions facilitates the demographical characterization of pebbles along the temporal dimension. Recent observations of these protostellar systems have revealed sizeable disks \citep{Tobin2020, Tychoniec2020, Sheehan2022} with enough mass to form planetary systems (see Sect.\,\ref{sec:opacity}) and dust grains several orders of magnitude larger than those in the pristine ISM \citep{Han2023, GuerraAlvarado2024}. Models suggest that dust can grow to pebble sizes in these early disks \citep[e.g.][]{Vorobyov2024}, but it remains unclear whether planetesimal formation via pebble accretion can occur under the strong turbulence driven by intense infall \citep{Huhn2025, Carrera2025}. In fact, in these embedded phases, infalling envelopes continue to supply gas and dust from the ISM \citep[e.g.][see also chapter by \citealt{Podio2026}]{Pineda2020, Garufi2022a}. The sensitivity of modern radio interferometers has made it possible to trace dust growth to pebble sizes even at envelope scales, providing constraints on the dust delivered to the disk. Evidence for dust growth to (sub-)millimetre sizes in infalling material comes from both NOEMA and ALMA \citep{Galametz2019, Cacciapuoti2025}. Recent ALMA observations have provided the first spatially resolved evidence for the presence of millimeter-sized grains along the dusty cavity walls of the Class I protobinary system L1551 IRS5 \citep{Sabatini25}. A remarkably low dust emissivity index ($\beta_{\rm mm}\sim1$) was measured in these regions, indicating grains that are significantly larger than those in the interstellar medium (e.g.\ \citealt{Mathis77, Dartois24}). This finding supports a scenario in which large dust grains -- grown in the inner protostellar disk -- are entrained into the surrounding medium by outflows and/or magnetohydrodynamic winds, enriching the envelope accreting onto young disks with these larger grains (see also \citealt{Giacalone19, Tsukamoto21, Bhandare2024, Cacciapuoti24b, Sabatini2024, Uchimura2025}). 

Even with only moderate resolution, a large-scale survey probing Class 0, I, and II could offer valuable insights into when and how pebbles grow and possibly fragment at various stages of disk evolution. In fact, the regions observable with SKAO span a range of ages (less than 1 to 5 Myr), and from a statistical perspective, their disk properties may reflect distinct underlying physical processes. An interesting scenario to test is whether pebbles are more abundant in regions at intermediate ages (2--3 Myr) than the expectation from a linear decrease of dust with time. This behaviour was studied observationally by \citet{Testi2022}, who found that the average dust mass in disks within Lupus and Chamaeleon is higher than expected from a simple monotonic decline with age — unlike the mass accretion rate, which consistently decreases over time. 
Theoretical models \citep{Bernabo2022} suggest that comparable patterns can emerge when planets forming at early stages (within 1-2 Myr) dynamically stir nearby planetesimals into eccentric and inclined orbits, initiating a cascade of collisions and fragmentation. These physical processes have been recently invoked to explain the origins of chondrules in the Solar System \citep{Sirono2025} and were found to constrain Jupiter's formation timescale to 1.8 Myr, consistently with the timescales proposed by \citet{Bernabo2022} to explain the dust observations by \cite{Testi2022}. 

These processes are expected to leave an imprint on the spatial distribution of cm-sized particles as well as on their total abundance \citep[see also chapter by][]{Wu2026}. Dust and pebbles produced by the collisions of planetesimals stirred by newly formed planets spatially favour the orbital regions inward of the planets, where the pericenters of their eccentric orbits are located. Conversely, the dust traps created by the planets prevent the primordial dust and pebbles from reaching these inner orbital regions. This spatial dichotomy between primordial and collisional pebbles would explain the anomalous dust distribution in HD163296 planet-forming disk \citep{Isella2016,Turrini2019} and is supported by the observed dichotomy in the dust properties in the inner and outer regions of this disk \citep{Guidi2022}. 

Finally, SKAO will enable a robust statistical comparison between disks around single stars and those in binary or higher-order multiple systems. By combining SKAO observations with data at shorter wavelengths, it will be possible to assess whether multiplicity alters the growth and dynamics of dust particles \citep{Zagaria2023, Alaguero2024, Cuello2025}. This population-level approach will be crucial to quantify how multiplicity affects disk mass, size, structure, and chemistry, and ultimately how it influences the pathways toward planet formation in different dynamical environments.

\subsection{Dust opacity and disk mass} \label{sec:opacity}
Solids are typically thought to contribute only about 1\% of the total mass in planet-forming disks. Nevertheless, dust opacity dominates over gas opacity, making solids the primary drivers of both the thermal and geometrical structure of the disk. Moreover, dust grains serve as the fundamental building blocks of planets, making accurate determination of the dust mass a central focus in the community. This focus has led to one of the major conundra of the field, that is the apparent shortfall of solid mass available for forming the observed exoplanetary systems \citep{Najita2014, Manara2018, Tychoniec2020}. This discrepancy suggests either a significant underestimation of dust mass or a need to revise our paradigm of planet formation, as most planets may possibly form at earlier stages of disk evolution.

Currently, the most direct method to estimate the disk dust mass $M_{\rm dust}$ involves converting the observed \mbox{(sub-)millimeter} continuum emission $F_\nu$ under the assumption of optically thin emission, according to:
\begin{equation}
       M_{\rm dust}=\frac{F_\nu d^2}{\kappa_\nu B_\nu(T_{\rm dust})}
\end{equation}
where $d$ is the source distance, $\kappa_\nu$ is the dust absorption opacity at that frequency and $B_\nu(T_{\rm dust})$ the Planck function at a given dust temperature $T_{\rm dust}$. This approach, however, is accompanied by significant uncertainties deriving from our limited knowledge of both dust properties and disk structure. On the one hand, as previously discussed, the high surface density in the inner disk regions {($\lesssim30$ au)} leads to optically thick emission even at ALMA frequencies. Within this framework, all dust mass estimates derived from ALMA continuum emission should be treated as lower limits. On the other hand, the connection between solid material and its observed emission depends on factors like dust opacity, albedo, and polarization -- none of which are yet well understood. Recent studies have shown that scattering, once considered negligible at millimeter wavelengths, plays a surprisingly significant role in shaping the observed continuum emission, directly impacting dust mass estimates \citep{Zhu2019, Liu2021}. This additional opacity contribution further complicates the interpretation of observed fluxes. A more robust, though less direct, approach would involve forward radiative transfer modeling, which accounts for optical depth effects, grain properties, and disk geometry to better constrain dust masses.

The particle size distribution plays a major role in shaping the emission. This distribution is typically described by a power law with an index $q$ that, at millimeter wavelengths, drives the falloff of the opacity with increasing maximum grain size. A common approach has long been to assume that $\kappa_\nu$ scales with frequency as $\nu^\beta$ \citep{Beckwith1990} with $\beta$ possibly measured from the slope of the (sub-)millimeter emission (see Sect.\,\ref{sec:spectral_pebbles}), although these indices are often affected by the optically thick contamination. In general, the choice of the maximum grain size leads to very different estimates of dust mass, and especially at (sub-)centimeter wavelengths where using different opacity models can result in extremely large variations of dust masses \citep[see e.g.][]{Birnstiel2018, Garufi2025}.

SKAO offers a promising avenue to address both the limitation of optically thick emission and the uncertainty relative to dust opacity. While the former will be immediate with the frequency range, the latter will require some effort on refining existing models of opacity or comparisons between the estimates derived from the SKAO optically thin emission and those obtained through alternative diagnostics of the total disk mass such as CO isotopologues \citep{Miotello2023} or deviations from a Keplerian motion for the disk rotation that are induced by the disk self-gravity \citep{Lodato2023}. Another promising method to directly constrain the total disk mass is through observations of the HD emission line at 112 $\mu$m \citep[see e.g.][]{Bergin2013} which enables a relatively straightforward conversion to the molecular abundance of H$_2$. Although no far-infrared space telescope is currently operational to detect this line, the approach remains highly compelling and may become feasible within the SKAO operational era, should future missions such as PRIMA (PRobe far-Infrared Mission for Astrophysics) restore access to this wavelength range. In any case, dedicated and comprehensive modeling efforts will be essential to reconcile different mass tracers and to constrain the gas-to-dust mass ratio and the other physical parameters of disks needed to determine their mass.

\subsection{Circumplanetary disks and protoplanets} \label{sec:cdp}

The detection of circumplanetary disks (CPDs) is among the most ambitious observational goals of the community. Their search and characterization with the SKAO presents both challenges and opportunities that are distinct from those faced by ALMA. While ALMA requires extremely high angular resolution ($\lesssim$50~mas) to disentangle CPD emission from the typically optically thick planet-forming disk \citep{Zhu2018}, SKAO may offer complementary avenues. In particular, the detectability of CPDs in the centimeter regime depends on the substantial presence of pebbles accumulated around the protoplanet \citep{Drazkowska2018, Long2022}. These pebbles, having already grown past the main barrier, falling at millimeter sizes in CPDs \citep{Zhu2018}, may emit more efficiently in these structures, potentially enhancing the contrast between CPDs and the surrounding disk material. 

Moreover, SKA could enable alternative detection strategies. Deviations from Keplerian rotation induced by the gravitational perturbation of an embedded planet can reveal localised velocity "kinks" in the gas kinematics \citep{PerezS2015, Pinte2019}. These features, observable through spectral line emission at angular resolutions of $\sim$0.1$''$ \citep{exoALMA_I}, also fall within the SKAO's expected capabilities. A key question is whether gas tracers emitting at SKA-Mid bands frequencies are bright enough to reconstruct the velocity field with sufficient fidelity \citep[see chapter by][]{Podio2026}. A potentially useful tracer of bright gas kinematics is H$_2$CO. Its line strength varies with temperature, and transitions in the SKA-Mid bands weaken when the gas is hot, but become stronger in cold environments ($\sim$10~K). This behaviour is particularly relevant in the mid-plane of some planet-forming disks, where temperatures drop below the CO freeze-out threshold ($\sim$20 K), leading to a layered structure with CO frozen out in the midplane and gas emission confined to the disk surfaces. {In this regard, the H$_2$CO 2(1,1)-2(1,2) line at 14.4885 GHz (with \mbox{E$_{\rm up}=$ 22.6 K}) would be the best possible proxy.} Colder tracers mapped with SKA could potentially open the window to study the gaseous component of the midplane and the kinematics of gas that is possibly perturbed by the presence of an embedded planet.

Furthermore, non-thermal cyclotron emission from planetary aurorae (see also chapters by \citealt{Vedantham2026} and \citealt{Kavanagh2026} or accretion shocks as well as free-free emission from jets and winds could also provide a detection pathway of CPD \citep{Gressel2013}, albeit within the lower-frequency bands of SKA, where angular resolution may limit spatial disentanglement from the central star. Finally, changes in the local chemical composition caused by the heating from a self-luminous planet \citep{Montesinos2015} may produce molecular abundance asymmetries detectable with radio interferometers \citep{Cleeves2015}, offering yet another indirect route for identifying embedded planetary companions with SKAO.

\subsection{Anomalous microwave emission} \label{sec:ame}

Anomalous Microwave Emission (AME) has emerged as a potential contributor to the centimeter-wave emission observed in some planet-forming disks, with tentative detections reported in systems such as V892 Tau \citep{Long2021}, HD97048, MWC297 \citep{Greaves2018}, CX Tau \citep{Curone2023}, and PDS 70 \citep{Liu2024}. AME is commonly attributed to electric dipole radiation from rotating nanometer-sized dust grains, also called spinning dust emission \citep{Draine1998}, although the identity of the carrier remains uncertain \citep{Draine2003ARA&A, Greaves2018}. While this emission is a natural outcome of dust models with abundant very small grains ($\lesssim$1~nm scales), alternative mechanisms, such as thermal magnetic dipole radiation from magnetic grains may also contribute \citep{Draine2003ARA&A}.

AME is expected to produce a distinct spectral feature: a broad bump in the spectral energy distribution peaking near 1 cm \citep[see][]{Curone2023}. This can resemble or contaminate other emission components in the same regime, particularly those from large grains or ionized gas. For further discussion on the role of free-free emission in disks, see the chapter by \citet{Guidi2026} and for an expanded discussion of AME see the chapter by \citet{Vidal2026}.

In the context of disk demographics, identifying AME offers a novel probe of the population of very small grains in circumstellar environments ---possibly nanodiamonds, as suggested by \citet{Greaves2018}. However, given the uncertainties surrounding both the emission mechanism and its manifestation in disks, robust confirmation requires high-sensitivity, simultaneous multiwavelength observations capable of disentangling AME from other cm-wave processes (e.g.\ free-free, thermal dust). The SKAO, with its broad frequency coverage and exceptional sensitivity in the cm regime, will be uniquely suited to address this challenge. A systematic search for AME across a diverse sample of disks would constrain the abundance, composition, and evolution of nanometric grains -- complementing the broader effort to characterize the solid content of disks at all scales. 

\subsection{Eruptive stars} \label{sec:eruptive_stars}
Eruptive young stellar objects (YSOs), typically classified as EXors or FUors, undergo episodic accretion bursts on timescales of months to years (EXors) or decades to centuries (FUors). These rare but dramatic events offer a unique laboratory to study early stellar evolution and the dynamical processes shaping planet-forming environments. These bursts, characterized by dramatic increases in stellar accretion rates by several orders of magnitude, were initially identified through abrupt optical brightening in their light curves \citep{Herbig1977, Audard2014}, and later confirmed through spectroscopic signatures of disk-related emission and absorption lines \citep[e.g.][]{Connelley2018}. This class of objects is particularly compelling for two reasons: first, they may hold the key to understanding how low-mass stars acquire their final mass, offering a resolution to the so-called luminosity problem in YSOs \citep{Kenyon1990, Dunham2010}; second, they provide rare, time-variable windows into the structure and chemistry of planet-forming disks as they respond to sudden changes in the physical conditions. The accretion variability in these objects can also be studied at different wavelengths, like near- and mid-IR \citep[e.g.][]{Zakri2022}.

Millimeter-wavelength surveys have shown that FUor disks are systematically more massive than those of EXors \citep{Cieza2018}. Moreover, EXor sources generally lack the prominent molecular outflows commonly observed in FUors \citep{Cieza2018}. Compared to typical Class I and Class II objects, the disks of eruptive stars (particularly FUors) are on average 3–4 times more massive and 2–5 times more compact in size \citep{Kospal2021}. A significant fraction of these disks, possibly exceeding 60\%, may be gravitationally unstable \citep{Kospal2021}.

Stellar multiplicity is another key factor in understanding the eruptive behavior of young stars. Observations show that a large fraction of stars form in multiple systems, and close stellar companions can significantly influence disk evolution and accretion variability \citep{Reipurth2014}. In particular, dynamical interactions within non-hierarchical or close binary systems --- such as periastron passages or orbital evolution --- can perturb circumstellar disks and trigger bursts of enhanced accretion onto the central protostar \citep{Bonnell1992, Reipurth2004}. High-resolution imaging studies have revealed a high incidence of close companions among eruptive stars, reinforcing this connection \citep{Zurlo2024}. Moreover, hydrodynamic simulations suggest that gravitational instabilities and disk fragmentation, which may be seeded or amplified by stellar encounters, can produce large-scale inflow events that resemble FUor-type accretion outbursts \citep{Vorobyov2015, Borchert2022a, Borchert2022b}. 

Episodic accretion enhances the brightness of disks at radio wavelengths, rendering distant systems --- even those several kiloparsecs (kpc) away --- temporarily as bright as nearby, passive planet-forming disks. These events can push snowlines outward, as in the remarkable case of V883 Ori where the water snowline was observed at tens of astronomical units \citep{Cieza2016, Tobin2023}. They also drive rich disk chemistry and vertical temperature inversions via viscous heating \citep{Lee2019}, and have been linked to large-scale instabilities. In one remarkable case, an accretion burst revealed spiral-driven fragmentation and the formation of a gravitationally bound clump around a low-mass protostar \citep{Weber2023b}. Despite the system's distance of over 2~kpc, the sheer power of the burst and its intricate connection to the large-scale environment make it an ideal target for high-resolution imaging. While the physical mechanisms behind these outbursts remain debated, they appear to link spatial scales ranging from hundreds of astronomical units down to sub-au regions.

Recent infrared surveys identify more than 700 FUor and EXor candidates. The majority of this sample are YSOs located at distances larger than 0.6 kpc and with a median value of 2.8 kpc \citep{Contreras2025}. Looking ahead, the SKAO will offer unprecedented sensitivity and resolution to probe thermal emission from these active disks, even at kiloparsec distances. Moreover, SKAO’s capabilities will make it possible to detect non-thermal emission associated with the most intense accretion episodes, providing critical constraints on the interplay between disk physics, chemistry, and accretion variability across a broad range of scales.

\section{Census and observability} \label{sec:sample}

In this section, we present predictions on the observability of actual targets, informed by our current understanding and available data. Our analysis begins with a census based on the ALMA surveys of planet-forming disks conducted over the past decade. We restrict our sample to sources within 200 pc and focus specifically on those with available ALMA observations, particularly targets for which measurements of continuum disk size and integrated flux have been reported.

\subsection{Sample and sensitivity}
The large Scorpius-Centaurus association {(including Lupus and Ophiuchus)} and Corona Australis are the ideal star-forming regions to be studied with SKAO given their distance (130--180 pc) and declination (-30\degree-- -40\degree). Therefore, we start building the sample from the targets in Upper Sco \citep{Barenfeld2016, Carpenter2025}, Lupus \citep{Ansdell2018}, Ophiuchus \citep[ODISEA;][]{Cieza2019, Cieza2021}, and Corona Australis \citep{Cazzoletti2019} with continuum disk extent and integrated flux tabulated. We also included sources from Chamaeleon \citep{Pascucci2016, Villenave2021} and Taurus \citep{Long2018} even though their declination represents a cut-off point in the declination range offered by SKAO. This sample already sums up to 360 sources. However, we included 30 additional sources of high scientific interest from exoALMA \citep{Teague2025} or from individual works \citep{Casassus2012, Andrews2016, Isella2016, Fedele2017, Francis2020}.

\begin{figure*}
    \centering
    \includegraphics[width=0.85\linewidth]{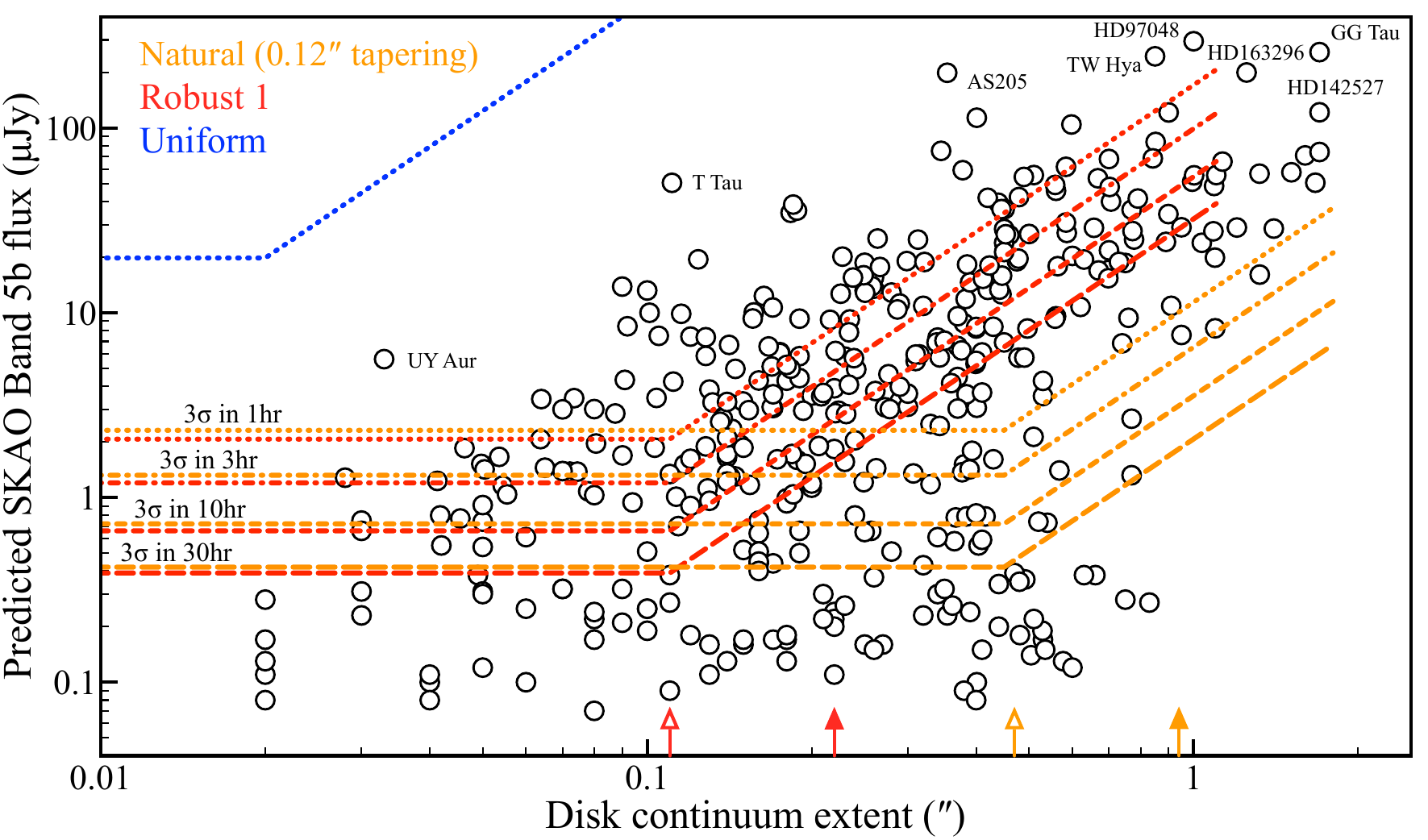}
    \caption{Predicted disk flux versus extent for the explored sample. The various lines indicate the 3$\sigma$ observability for different disk extent, integration time, and weighting parameters (red lines for Briggs weighting with Robust 1, orange for a tapered Natural weight, and blue for Uniform for which only the one~hour case is shown). Empty and filled arrows indicate the extent over which disks are mildly and highly resolved (see text) for the Briggs and tapered Natural weighting cases.}
    \label{fig:observability}
\end{figure*}

The prediction on each source brightness is based on the recent VLA study by \citet{Garufi2025}. In this work, an average spectral index of 2.3 between 1.3 mm and 3 mm was determined for a large set of ordinary disks in Taurus (see the example of Fig.\,\ref{fig:sed_component}). After removing the free-free emission, the dust average spectral index was found to be 2.8. In absence of any better observational constraint, here we assume that the entire sample in question exhibits the same behaviour and thus converts the measured ALMA fluxes into a predicted SKAO flux in Band 5b ($\sim3$ cm). By doing so, the observed fluxes spanning from 0.2 mJy to 1050 mJy with an average of 50 mJy turns into fluxes spanning from 0.07 $\mu$Jy to 295 $\mu$Jy with an average of 13 $\mu$Jy. Regarding the disk extent in Band 5b, few observational constraints have been obtained to date \citep[see e.g.\ TW Hya with VLA by][]{Wilner2005}, making this one of the primary objectives of SKAO (see Sect.\,\ref{sec:spatial_pebbles}). In this work, we adopt the disk extent inferred from ALMA observations as an upper limit and explore scenarios with smaller extents (see below). The predicted disk fluxes and extents for the sample under investigation are shown in Fig.\,\ref{fig:observability}.

To evaluate the observability of the sample we use the SKAO Sensitivity Calculator version 1.5.0. {This exercise is performed adopting the Array Assembly 4 (AA4) with the 15-m antennas only. However, the array is planned to operate in conjunction with the 64 MeerKAT antennas (13.5 m diameter), resulting in an improved sensitivity of an approximate factor 1.2 that is not considered here.} We first investigated the impact of source position in the sky and found only modest variations in sensitivity (up to approximately 50\% between targets in Lupus and Chamaeleon). Based on this, we focused our analysis on a representative case of a target in Lupus at an altitude of 30$\degree$, which falls near the midpoint of the sensitivity range. We then compared the sensitivity and resolution achieved with different weighting schemes against the requirements set by the predicted fluxes and spatial extents of our sample. The first key result is that Uniform weighting proves unsuitable for our observations, as it yields a null number of detectable disks even at long exposures. Consequently, our analysis focuses on the range between Natural weighting and Briggs weighting with Robust = 0. These weighting choices clearly illustrate the trade-off between sensitivity and resolution, enhancing our ability to either detect or resolve the disks under study. An illustrative example is provided in Fig.~\ref{fig:observability} where we show the 3$\sigma$ observability with Briggs weighting and Robust 1 as well as with the Natural weighting after applying a 0.12$"$ tapering\footnote{Natural weighting with no tapering yields a non-gaussian beam whereas, with the smallest available tapering (0.03$"$), the resulting sensitivity and resolution are very similar to the Briggs weighting with Robust 1.}. The former weighting bears a 0.11$"$ resolution and thus allows us to resolve a much larger number of disks but prevents the detection of a large reservoir of disks with a putative extent larger than 0.3$"$. Conversely, the tapered Natural weighting enables the detection of a larger number of targets but restricts the resolved characterization to a handful of objects.

\begin{figure*}
    \centering
    \includegraphics[width=1.0\linewidth]{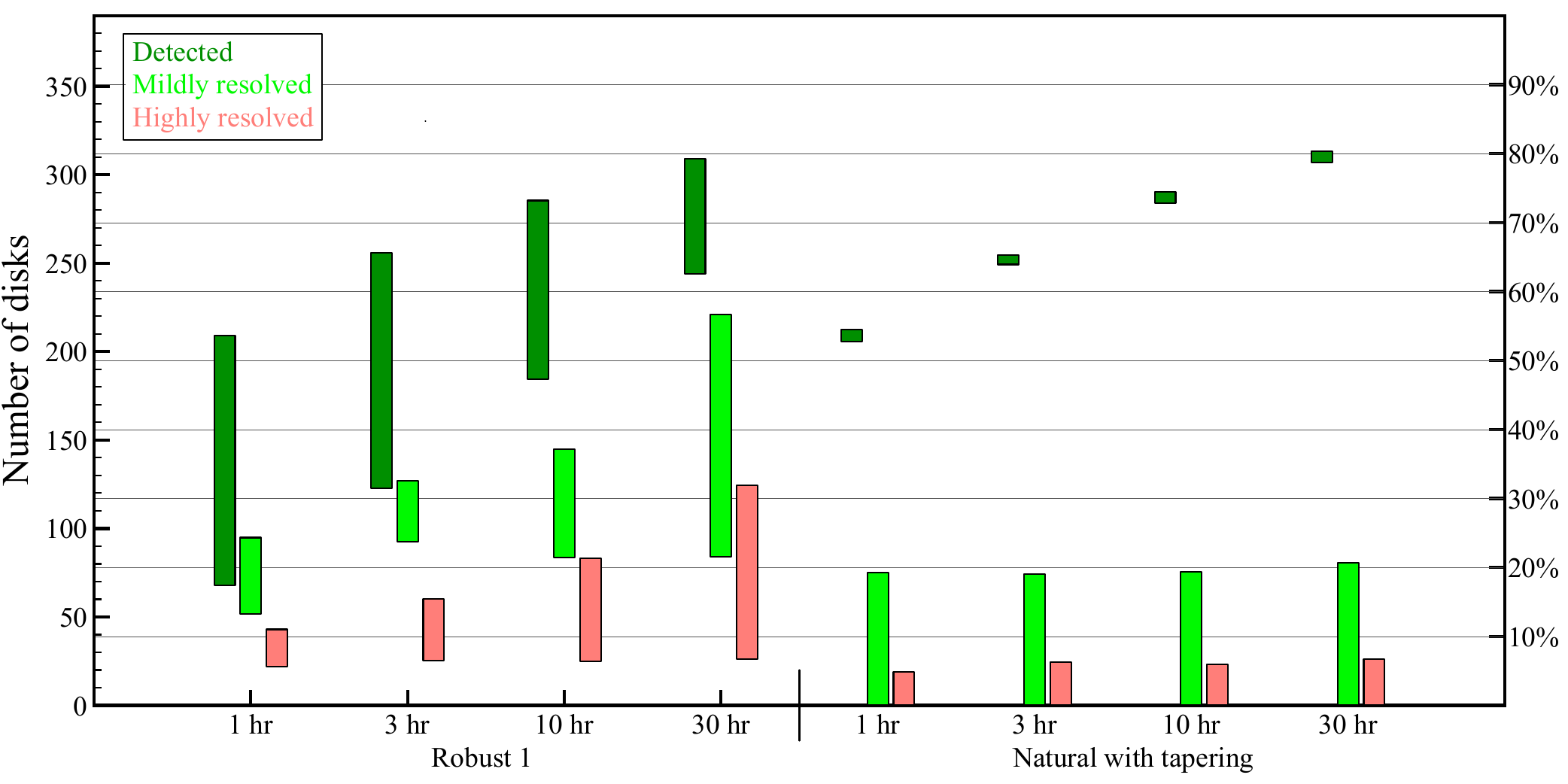}
    \caption{Number of disks detected and resolved with different weighting schemes and integration times. The bars indicate the range of detected or resolved objects depending on the disk extent in Band 5b (from 25\% to 100\% of the continuum extent measured by ALMA).}
    \label{fig:detected_disks}
\end{figure*}

\subsection{Predicted detection statistics}

To obtain more quantitative estimates that also account for the range of possible disk extents, in Fig.\,\ref{fig:detected_disks} we count the number of targets detected and resolved with the two weighting schemes shown in Fig.\,\ref{fig:observability}. We consider two extreme cases of disk extent: a radius as large as the ALMA radius ($r=1$) and a radius of one fourth ($r=0.25$). We consider a disk resolved only if the extent is larger than the angular resolution and highly resolved if it is at least twice as much (thus with the disk imaged in four resolution elements). The results of this census can therefore be summarized as follows:

\begin{itemize}
    \item 80 of our 390 disks (20\%) cannot be detected in less than 30 hours of integration regardless of their size. Thus, a significant fraction of disks would require extremely long exposures to even be detected.
    \item The number of detectable disks depends strongly on their extent when adopting a \mbox{Robust = 1} weighting, spanning from barely 68 (less than 20\%) for a large extent ($r=1$) in one hour of observation to more than 260 (60\%) for small disk extents ($r<0.5$) and at least three hours of observations. In case of tapered, Natural weighting, this number only depends on the observation time and in a mild manner (spanning from 50\% in only one hour to 80\% in 30 hours).
    \item The number of resolvable disks strongly varies with the observing time. Around 80 of our disks (20\%) would be resolved in barely one hour with Robust 1 while more than 10 hours are needed to resolve twice as many.
    \item 25 to 40 disks (around 10\%) would be significantly resolved in only one hour with Robust 1. This number can in principle reach 100 (25\%) for exposures longer than 10 hours.
\end{itemize}

It is clear that using Natural weighting with modest tapering ensures the detection of a statistically significant number of disks -- over 200 -- in just one hour of observation, regardless of the disk size at centimeter wavelengths. However, such detections alone are insufficient to disentangle the dust emission from contributions by ionized gas. Therefore, these observations are meaningful only when combined with spectral studies (see Sect.\,\ref{sec:spectral_pebbles}). Imaging with Briggs weighting at \mbox{Robust = 1}, on the other hand, would allow us to spatially separate star-related processes from disk emission in up to 100 disks with only three hours of integration. Finally, highly resolved imaging of the disk emission would be achieved for 25 sources, regardless of their size relative to ALMA (down to at least one-quarter of it) even with relatively short integration times.

\subsection{Special targets}

Within the broader sample discussed above, certain sources stand out as particularly compelling targets for dedicated observations. These objects have been the focus of long-standing ALMA and high-contrast imaging studies and have drawn significant attention from the community for a variety of reasons. Some of these sources are highlighted in Fig.\,\ref{fig:observability}. With few exceptions, their disks are notably bright at millimeter wavelengths and are thus expected to be similarly bright at centimeter wavelengths -- making them likely early targets for SKA. This continues the inevitable observational bias of first focusing on the most accessible systems.

The relationship between disk brightness and spatial extent, illustrated in Fig.\,\ref{fig:observability}, is critical for planning observation strategies and data reduction. For instance, compact but bright disks such as those around UY Aur, {IRAS 12496-7650, HD98800}, and AS 205 may benefit from high angular resolution, favoring imaging schemes like Briggs weighting with \mbox{Robust = 1}. Conversely, extended disks such as those of {TW Hya, HD97048,} HD163296, and HD142527, are better suited for tapered Natural weighting to enhance surface brightness sensitivity -- unless long integration times are carried out.

These targets span a broad range of scientific cases, from disk interactions in wide binaries (e.g.\ {S CrA}, T Tau) and environmental effects (e.g.\ {T CrA}, SU Aur), to the structure of circumbinary disks (e.g.\ {WW Cha, AK Sco}, HD142527). A major scientific goal will be the characterization of substructures linked to planet formation. While confirmed protoplanets are currently limited to only PDS 70 {and WISPIT 2 \citep{Keppler2018, vancappelleven2025}}, it is realistic to expect that more will be identified by the time SKA becomes operational. The role of these systems in tracing planet-disk interactions is discussed in detail in the chapter by \citet{Wu2026}.

\subsection{Surveys of star-forming regions} \label{sec:sample_deepfield}

\begin{figure*}
    \centering
    \centering\includegraphics[width=0.8\linewidth]{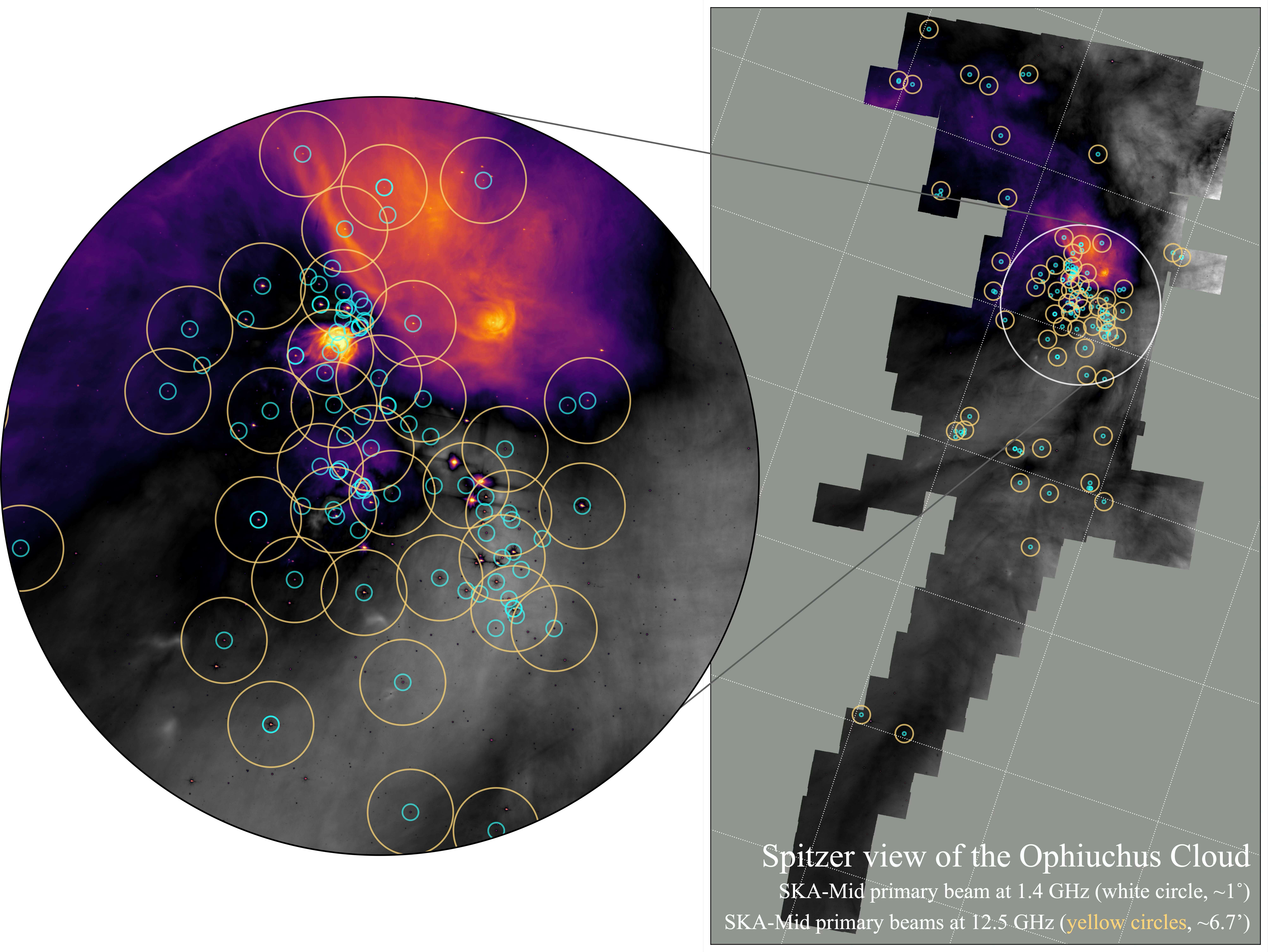}
    \caption{{\em Spitzer} map of the Ophiuchus molecular cloud from the 'Cores to Disks' Legacy Project. Cyan circles indicate the locations of planet-forming disks detected in the ODISEA survey \citep{Cieza2019}. The large white circle shows the SKA-Mid primary beam at 1.4\,GHz ($\sim$1$~\deg$), while the yellow circles mark the primary beams at 12.5\,GHz (6.7'). Field-of-view sizes correspond to the primary beam of a 15-m dish. Covering the entire disk population requires $\sim$60 pointings at 12.5\,GHz, though fewer are needed at longer wavelengths as the primary beam increases. The inset highlights one 1.4\,GHz primary beam centered on the most crowded region of the cloud.}
    \label{fig:pointings}
\end{figure*}

The wide field of view of the SKA will enable population-level studies of entire star-forming regions with only a modest number of pointings. As shown in Fig.~\ref{fig:pointings}, the Ophiuchus molecular cloud, one of the nearest and best-studied low-mass star-forming regions, can be efficiently covered with SKA-Mid thanks to its large primary beam {\citep[see][]{Hoare2015}. In the 10--1 GHz regime,} the FoV of a 15-m dish spans {from several arcminutes to} nearly one degree, which is comparable to the angular extent of the cloud itself. This makes it possible to simultaneously capture dozens of YSOs and their disks within a single primary beam, in contrast with millimeter and infrared facilities where mosaicking requires hundreds of pointings to achieve comparable coverage. The ability to probe entire disk populations in their native environments will be crucial for connecting disk demographics with cloud structure and large-scale star formation processes.  

To provide a first-order estimate of how many SKA pointings are required to cover the Ophiuchus disk population, we employed a simple beam-placement algorithm. Starting from the source list of ODISEA disks \citep{Cieza2019}, a beam is placed on a given disk and a KD-tree is used to identify all neighboring disks within the beam. These are then removed from the tree, and the procedure is repeated until no sources remain. While this approach is not optimized, it yields a practical tiling of the cloud with reasonable efficiency. Applying this method at 12.5\,GHz, where the primary beam is $\sim$6.7$'$, we find that approximately 60 pointings are sufficient to cover the full set of ODISEA disks. This demonstrates that SKA observations will be able to probe complete disk populations across an entire star-forming region with a relatively small investment of observing time, offering a transformative perspective on the interplay between environment and disk evolution.  

Beyond Ophiuchus, the SKA will enable comprehensive demographic studies of disks across nearby star-forming regions accessible from the southern hemisphere, such as Lupus, Upper Sco, and Orion, as well as the more northern Taurus and more southern Chamaeleon complexes. As illustrated in Fig.~\ref{fig:pointings} for Ophiuchus, the wide FoV of SKA makes it possible to design efficient pointing strategies that encompass entire disk populations, a capability that can be generalized to many of these regions. With its sensitivity and frequency coverage, the SKA will allow us to constrain the contents of millimeter- and centimeter-sized dust grains with unprecedented accuracy, with the cm continuum providing critical leverage on the larger solids. In addition, the central star's activity, revealed through free-free emission (see chapter by \citealt{Guidi2026}), will be simultaneously characterized. By surveying multiple regions with diverse ages and environments, SKA observations will allow us to disentangle evolutionary trends from environmental effects. {Importantly, unlike studies that rely on the highest angular resolution to resolve fine disk substructures, this type of population-level analysis can be effectively carried out already with early SKA capabilities (in AA$*$), without requiring the full AA4 configuration.} Thanks to its survey capabilities, the SKA-Mid will deliver a legacy dataset of long-lasting value, enabling rigorous tests of theoretical models of disk evolution and planet formation across diverse environments.

\section{Conclusions}
Over the past decade, the field of planet formation has been significantly advanced by the unprecedented capabilities of ground-based high-contrast imaging and of ALMA. Resolved imaging of planet-forming disks has become routine, and the resulting large-scale observational census -- now encompassing several hundred individual disks -- has spurred major theoretical progress. Despite these advances, a key challenge remains: the limited time available for dust grains to overcome the meter-size barrier within the classical core accretion framework. Alternative pathways, such as pebble accretion, offer promising solutions but require detailed observational characterization of centimeter-sized grains -- a task that remains out of reach with current observational facilities.

The SKAO will be a transformative facility for characterizing the demographics of planet-forming disks. Its sensitivity and angular resolution at centimeter wavelengths will allow the first statistically significant exploration of pebble-sized grains across a wide variety of disks, overcoming key limitations of (sub-)millimeter observations with ALMA, such as optical depth effects and limited sensitivity to larger grains. This chapter presents the motivations, methodology, and predictive analysis for SKAO observations of nearby star-forming regions, demonstrating that up to 300 disks will be detectable in reasonable telescope times and up to a few tens of them spatially resolved, even with modest integration times.

Several science cases are developed in this chapter. The primary focus is the spatial, spectral, and temporal characterization of pebbles in planet-forming disks from nearby star-forming regions, which will enable the community to understand how, where, and when the most critical stages of dust growth toward planet formation occur. In addition, we outline several complementary science cases that SKAO will address. These range from improving estimates of disk dust mass and opacity through optically thin emission, to detecting and characterizing circumplanetary disks, and exploring lesser-known processes such as anomalous microwave emission or more distant targets like eruptive stars at kiloparsec-scale distances. Taken together, these efforts will make the {extensive SKAO survey of planet-forming disks} a cornerstone for next-generation planet formation models and, alongside the E-ELT, a leading driver of observational planetary science.

{Looking beyond the AA4 design, several upgrades would significantly enhance the impact of SKAO for planet-forming disk studies. First, extending the frequency coverage toward higher frequencies (e.g., a potential Band 6 covering $\sim$15--50 GHz) would provide a critical bridge between SKA and ALMA, enabling observations of dust emission closer to the peak of the spectral energy distribution while remaining in a largely optically thin regime. Second, longer baselines -- including the possibility of SKA–VLBI capabilities -- would push the angular resolution to sub-au scales in nearby regions, opening the door to resolving the innermost disk structure and potentially circumplanetary environments. Finally, increased collecting area (or improved sensitivity) would directly translate into access to the bulk of the disk population currently below detection thresholds, enabling true demographic studies across stellar mass, age, and environment. Together, these developments would transform SKAO from a facility primarily probing the cm-sized grain population into a fully multi-scale, multi-grain tracer of disk evolution.}

\section*{Acknowledgements}
\textit{We thank the anonymous referee for the constructive and insightful comments. The research activities described in this paper were carried out with contribution of the Next Generation EU funds within the National Recovery and Resilience Plan (PNRR), Mission 4 - Education and Research, Component 2 - From Research to Business (M4C2), Investment Line 3.1 - Strengthening and creation of Research Infrastructures, Project IR0000034 - "STILES - Strengthening the Italian Leadership in ELT and SKA". S.P. acknowledges support from FONDECYT 1231663, ANID--Millennium Science Initiative Program NCN2024$\_$001 and ANID FIUF137139-USACH. JDI acknowledges support from an STFC Ernest Rutherford Fellowship (ST/W004119/1). Y.W. acknowledges the EACOA Fellowship awarded by the East Asia Core Observatories Association. This research was supported by the funding from the National SKA Program of China under Grant No.\,2025SKA0120100. We acknowledge the use of artificial intelligence for the purpose of improving the text's readability. Part of this research was by MN carried out at the Jet Propulsion Laboratory, California Institute of Technology, under a contract with the National Aeronautics and Space Administration (80NM0018D0004).}

\bibliographystyle{abbrvnat-maxbibnames4}
\bibliography{chapter} 

\end{document}